\documentclass{elsart3}
\usepackage{amsmath}
\usepackage{amssymb}
 
\newcommand{\be}{  \begin{eqnarray} }
\newcommand{\ee}{  \end{eqnarray} }

\def\spose#1{\hbox to 0pt{#1\hss}}
\def\lta{\mathrel{\spose{\lower 3pt\hbox{$\mathchar"218$}}
     \raise 2.0pt\hbox{$\mathchar"13C$}}}
\def\gta{\mathrel{\spose{\lower 3pt\hbox{$\mathchar"218$}}
     \raise 2.0pt\hbox{$\mathchar"13E$}}}

\begin{document}

\begin{frontmatter}

\title{Turbulent Comptonization in Relativistic Accretion Disks}
\author[Princeton]{Aristotle Socrates\thanksref{HST}},
\author[UCSB]{Shane W. Davis}, and
\author[UCSB]{Omer Blaes}
\address[Princeton]{Department of Astrophysical Sciences, 
Princeton University, 
Peyton Hall-Ivy Lane, Princeton, NJ 08544}
\address[UCSB]{Department of Physics, University of California, 
Santa Barbara, CA, 93106} 
\thanks[HST]{Hubble Fellow: socrates@astro.princeton.edu}

\begin{abstract}
Turbulent Comptonization, a potentially important damping and
radiation mechanism in relativistic accretion flows, is discussed.
Particular emphasis is placed on the physical basis, relative
importance, and thermodynamics of turbulent Comptonization.  The
effects of metal-absorption opacity on the spectral component
resulting from turbulent Comptonization is considered as well.

\end{abstract}

\end{frontmatter}

\section{Overview of Thermal and Bulk Comptonization}\label{s:1}

If a beam of photons interacts with a scattering
layer consisting of relatively hot electrons such that $k_BT_e>>h\nu$,
the input seed photons gain energy while conserving their overall
number, a process known as Compton up-scattering or ``Comptonization''
\cite{ral79}.  In the non-relativistic limit, the shift
in photon energy per scattering is small, so that $\Delta
E/E\sim k_BT_e/ m_ec^2$.  The rare photons that 
remain in the scattering layer the longest are subsequently boosted 
to energies limited by the electron temperature  -- a process
that is statistically identical to the classical theory 
of cosmic ray acceleration \cite{fer49} -- and the resulting 
spectral energy
distribution resembles a power law.
     
The fractional shift in energy per scattering event is proportional to
the mean square electron velocity i.e., $\Delta E/E\sim
\left<v^2_e\right>/c^2$ where $\left<\right>$ is taken over the
electron distribution function.  Also, if the scattering layer moves
in bulk with respect to the observer, then the radiation will
experience a Doppler shift such that $\Delta E/E\sim v/c$, an effect
often referred to in cosmological contexts as the ``kinetic Sunyaev-
Zeldovich effect.''  However, if the scattering layer is turbulent and
stationary with respect to the observer, then the Doppler shift term
that is $\propto v/c$ vanishes despite the presence of large scale
bulk motions.  It follows that the contribution to Comptonization
from turbulence leads to an amplification $\Delta E/E\sim v^2_{\rm
T}/c^2$, where $v_{\rm T}$ is the turbulent velocity.  This effect,
which we refer to as ``turbulent Comptonization,'' deforms the spectra
of soft photons in a manner indistinguishable from that of thermal
Comptonization \cite{pal97},\cite{tho94}.\footnote{In the language of
cosmology, turbulent Comptonization might be referred to as the
``second-order kinetic Sunyaev-Zeldovich effect.''}

In this note, we discuss certain aspects of turbulent Comptonization 
in the context of relativistic accretion disk theory.  Many of the 
results summarized here can be found in \cite{sdb04}.  

\section{Presence of Turbulent Comptonization in 
Relativistic Accretion Flows}\label{s:2}

Turbulent Comptonization dominates over thermal Comptonization
when $v^2_T>v^2_{th}$, where $v_{th}$ is the electron thermal 
velocity.  In what follows, we briefly outline why it is possible to  
satisfy this condition in relativistic accretion flows.  
  
Consider a thin accretion disk extending down to the innermost
stable circular orbit around a rapidly rotating black hole
\cite{nat73}, \cite{ss73}.  The orbital velocity near its inner-edge
approaches the speed of light.  For accretion rates close to the
Eddington limit, the inner-edge of the disk becomes marginally thick
i.e., the disk scale height $H$ is comparable to its radius $R$ from
the black hole.  As a result, the sound speed of the disk $c_s$ near
the central object approaches the speed of light since the thermal
content of the disk is close to the virial value.  In this
parameter regime, thin relativistic accretion disks are radiation
pressure dominated, potentially allowing for the radiation sound speed
to exceed the electron thermal velocity.  By assuming that the
accretion stress $\tau_{R\phi}$ both scales with the total pressure
and is turbulent in nature \cite{bah91}, we conclude that {\it in
principle}, the turbulent velocity $v_{\rm T}>v_{th}$ close to the
central gravitating object.  Therefore, it is possible for turbulent
Comptonization, rather than thermal Comptonization to be the radiation
mechanism responsible for the radiative release of gravitational
energy.  Note that the above arguments are {\it independent of black
hole (or neutron star) mass}.

The radiative emission of both Galactic X-ray binaries (XRBs) and active
galactic nuclei (AGNs) are most commonly modeled with a multi-temperature
disk component, accompanied by a Comptonizing corona.  Generally
speaking, the physics of two-phased systems that are composed of a
relatively cool, dense, optically thick, turbulent layer which
releases a fraction of its binding energy in a hot diffuse optically
thin corona is not well-understood.  A good example of a system
that illustrates our lack of predictive power is the solar
chromosphere and corona.  In other words, there remains a great deal
of uncertainty in identifying the physical mechanisms responsible for
converting dissipated gravitational binding energy into radiative
power.

Provided that an adequate supply of soft seed photons is available,
the spectral component arising from turbulent Comptonization does not
suffer from the same level of theoretical uncertainty as one modeled
from a multi-temperature disk + corona.  Specifically, the physical
mechanism responsible for the accretion is also directly
responsible for at least a portion of the radiative energy release.

\section{Transfer Effects and the {\it y}-parameter}    

At this preliminary level of accuracy, we model accretion disk
turbulence as a non-linear cascade of energy down to small scales,
where the outer driving scale of the turbulence is occupied and
powered by relatively large-scale modes driven by the
magnetorotational instability (MRI) \cite{bah91}.  Radiation pressure
supported thin accretion flow onto relativistic objects are optically
thick to Thomson scattering.  In a one-zone height-averaged disk
model, the photon mean free path $\lambda_p$ is likely to be small in
comparison to the outer scale of an MRI cascade.  If this is the case,
photons cannot ``sample'' the velocities of the most powerful eddies.

If, for example, accretion disk turbulence follows an isotropic 
Kolmogorov scaling, the velocity amplitude of the turbulence scales
as $v_{\rm T}\left(\lambda\right)\sim v_{\rm T}\left(\lambda_0\right)\left(
{\lambda}/{\lambda_0}\right)^{1/3}$, implying that
\be
T_{\rm T}\left(\lambda\right)\sim T_{\rm T}\left(\lambda_0\right)\left(
\frac{\lambda}{\lambda_0}\right)^{2/3}.
\ee
Here, $\lambda_0$ represents the outer scale of the turbulence and 
$T_{\rm T}\equiv 3 m_e v^2_{\rm T}/2k_B$ can be thought of as a 
turbulent electron temperature.  On scales where $\lambda=\lambda_p$, 
seed photons can efficiently extract energy from the turbulence 
as they are being boosted by it.   

The arguments presented in \S\ref{s:1} suggest that the
spectral deformation arising from turbulent Comptonization is
identical to that of thermal Comptonization.  For example, in the 
unsaturated case, the radiation spectrum due to a single 
homogeneous scattering layer is represented by a cut-off temperature, 
$y$-parameter, and normalization.  In \S\ref{s:2}, we showed that the 
outer scale turbulent velocity, and therefore temperature, of thin 
accretion flows is independent of the mass of the central object.  
Likewise, the turbulent $y$-parameter on the outer scale is 
given by  
\be
y_{\rm T}\sim \frac{4k_B\,T_{\rm T}\left(\lambda_0\right)}{m_ec^2}\,
\tau^2,
\label{e:y}
\ee        
which for thin disks, is {\it independent of mass and radius}
near the inner-edge of the disk \cite{sdb04}, where a great fraction 
of the binding energy is radiated. 

Perhaps it is not surprising that the Compton temperature of XRBs
and AGNs are comparable in value, despite the enormous range in central
mass.  That is, Compton temperatures typical to both systems hover
close to $\sim 100\,{\rm keV}$ -- the electron virial temperature a
few gravitational radii away from the surface of the black hole.  The
Compton $y$-parameter for XRBs and AGN are quite similar as well,
implying that the fraction of energy dissipated per unit area above
the depth of formation of the cool optically thick component is
comparable in the two cases.  If turbulent Comptonization is responsible
for the broad band X-ray emission of relativistic accretion flows,
then the arguments surrounding eq. (\ref{e:y}) might explain why the
spectral shape is similar in the two classes of sources.

\section{Compton Cooling of Accretion Disk Turbulence}

An important check of the internal consistency of our previous
arguments is to make sure that the rate of energy injected into the 
turbulent cascade is equal to the rate of energy gain by the radiation 
field.  A version of the Kompaneets equation that 
takes into account bulk motions \cite{Hu94},\cite{tho94},\cite{pal97}
allows one to estimate the turbulent Compton cooling rate
\be
t^{-1}_{\rm C,T}\sim \sigma_{es}n_ec\frac{v^2_{\rm T}}{3c^2}=
\sigma_{es}n_ec\frac{k_BT_{\rm T}}{m_ec^2}
\ee
where $\sigma_{es}$ and $n_e$ is the electron scattering cross section
and the electron number density, respectively.  The above expression
can also be interpreted as the approximate rate at which turbulence is
``cooled'' or damped by the radiation field.  For thin disks, Socrates
et al. \cite{sdb04} point out that
\be
t^{-1}_{\rm C,T}\sim\frac{\tau}{c}\frac{\alpha P}{H\rho}\sim
\frac{\alpha\tau}{c}\Omega^2H\sim\Omega
\label{e:cooling}
\ee
where $\Omega$, $\alpha$, $P$, and $\rho$ is the Keplerian frequency, the 
thin disk viscosity parameter, mid-plane pressure, and density, 
respectively.

If accretion disk turbulence is driven by the MRI, then the rate at
which energy is injected into the turbulent cascade i.e., the
eddy-turnover time $\sim\Omega$.  Remarkably, the rate of energy
injection into the random turbulent motions is comparable to the rate
at which the radiation field extracts energy from them.  An
implication that stems from eq. (\ref{e:cooling}) is that turbulent
Comptonization may serve as an important source of dissipation for
accretion disk turbulence, while also functioning as a radiation
mechanism which mediates the release of binding energy.

\section{Ionization Balance and Metal Edge Opacity}

In order to amplify seed photons, Comptonizing turbulence must be
bound within a Thomson scattering layer, where the effective optical
depth to absorption is less than unity.  Near the inner-edge, the disk
is likely to be optically thin, but effectively thick to free-free
emission.  In this region, the site of formation is close to the
mid-plane, implying that continuum absorption is negligible in the
above adjacent scattering layer, for a significant number of
scattering depths \cite{sdb04}.

Though it is straightforward to argue that radiation pressure
supported accretion disks are effectively thin to free-free 
absorption down to significant depth, the notion that the same holds
true for sources of bound-free absorption is not 
so easily grasped.  Again, the key issue circles around whether or 
not it is more probable for a seed photon to be up-scattered to 
X-ray energies or to be absorbed by, in this case, a C,N,O, or Fe 
atom.  To answer this question, we estimate the optical depth 
to absorption from metal edges.  

In order to obtain a rough idea of how important metal edges are,
we work with a one-zone model.  Also, for the sake of simplicity, 
we consider the case of saturated turbulent Comptonization, where
the energy of all of the disk photons are roughly given by 
$k_BT_{\rm T}\left({\lambda_p}\right)$.  We further assume that every
bound electron lies in a hydrogenic state of some fiducial ion of 
abundance $A_Z$.    


The above simplifications allow for an estimation of the hydrogen-like
metal ionization fraction by equating the photo-ionization rate to the
recombination rate i.e.,
\be
n_i\int^{\infty}_{\nu_0}d\nu\frac{E_{\nu}\sigma_{\nu}}{h\nu}\,c\simeq
n_in_{\gamma,I}\sigma_0c=n_en_{i+1}\alpha_R(T),
\ee
where $n_i$, $n_{i+1}$, $n_e$, $\sigma_0$, and $\alpha_R$ is the 
number density of bound ions, the number density of stripped ions, 
the number density of liberated electrons, the photo-ionization 
cross section at the threshold photon frequency $\nu_0$, and the 
recombination rate, respectively.  Also, the number density of 
ionizing photons near the midplane $n_{\gamma,I}$ is roughly given by
$n_{\gamma,I}\sim P_{r}/k_BT_T$, where $P_{r}$ is the radiation 
pressure at the disk midplane.

The ionization fraction $\chi_i\equiv n_i/n_{i+1}$ is given by
\be
\chi_i\simeq\frac{n_e\,\alpha_R(T)}{n_{\gamma,I}\,\sigma_0\,c}
\sim \frac{T_T}{T_d}\frac{P_g}{P_r}\frac{\alpha_R(T)}{\sigma_0\,c}.
\ee     
Here, $P_g$ is the midplane gas pressure.  This allows us to compare the 
optical depth resulting from bound-free absorption 
$\tau_{abs}$ to that of Thomson scattering $\tau_{es}$
\be
\frac{\tau_{abs}}{\tau_{es}}\sim\frac{A_Z\chi_i\sigma_0}{\sigma_{es}}
\sim A_Z \frac{T_T}{T_d}\frac{P_g}{P_r}\frac{\alpha_R(T)}{\sigma_{es}\,c}
\sim 3\times 10^{-3}
\ee  
for $A_Z=10^{-3}$, $T_T/T_d=10^2$, $P_g/P_r=10^{-5}$, and $\alpha_{R}
\left(T\right)=10^{-10}$.  Note that $\tau_{es}\sim\left(
\alpha\,{\dot m}\right)^{-1}$ for thin disks, implying 
that $\tau_{eff}\sim \sqrt{\tau_{abs}\tau_{es}}\lesssim 1$ if
$\alpha \sim 0.1$ and ${\dot m}\lesssim {\dot m}_{Edd}$.  Thus, 
it seems plausible that photons amplified by Comptonizing turbulence can 
survive long enough to escape, before being absorbed by the 
metal content of the disk.        



\section{Discussion}

The mechanisms that result in the randomization of gravitational
binding energy, transport of angular momentum, and the conversion of
binding energy into heat and radiation must be
simultaneously understood if a {\it predictive} theory of accretion is
desired.  It is widely believed that MRI turbulence is responsible for
the randomization of gravitational binding energy and the transport of
angular momentum\cite{bah91}.  Also, sophisticated techniques used to
model the spectra of relativistic accretion flows fit the data quite
well.  Nevertheless, there is currently no accepted method of
connecting the underlying MHD to the observable radiation spectrum,
implying that accretion theory suffers from an overall lack of
predictive power.  We argue that turbulent Comptonization alleviates
some of these inadequacies even if it is not the dominant mechanism of
radiative energy release, since it allows observers to peer into the
inner workings of the turbulence which is thought to be the prime
mover of the accretion process.

\section{Acknowledgements}
We thank J. Krolik and J. Poutanen for raising the issue 
of photo-electric absorption.

{}


\begin{thebibliography}{}

\bibitem{bah91} Balbus, S.A., \& Hawley, J.F. 1991,
ApJ, 376, 214

\bibitem{bel99} Beloborodov, A. 1999, ApJ, 510,
123

\bibitem{fer49} Fermi, E. 1949, PhRv, 75, 1169 

\bibitem{Hu94} Hu, W., Scott, D., \& Silk, J.1994, PhRvD, 
49, 2, 648

\bibitem{nat73} Novikov, I.D. \& Thorne, K.S. 1973, in
Black Holes, eds. C. De Witt and B. De Witt (New York: Gordon \& 
Breach) p. 343

\bibitem{pal97} Psaltis, D., \& Lamb, F.K. 1997, ApJ, 488, 881

\bibitem{ral79} Rybicki, G.B., \& Lightman, A.P. 1979, Radiative 
Processes in Astrophysics (New York: Wiley)

\bibitem{sdb04} Socrates, A., Davis, S.W., \& Blaes, O. 2004, ApJ, 
601, 405

\bibitem{ss73} Shakura, N.I. \& Sunyaev, R. A. 1973, A\&A, 24, 337

\bibitem{tho94} Thompson, C. 1994, MNRAS, 270, 480

\end{thebibliography}
\end{document}